\documentstyle[11pt,aaspp4]{article}      
 
\lefthead{Sarzi et al.} 
\righthead{SuperMassive Black Holes in Bulges} 
 
\begin{document} 
 
\title{SuperMassive Black Holes in Bulges\footnotemark}  
\footnotetext{Based on observations with the {\sl Hubble Space
Telescope} obtained at STScI, which is operated by AURA, Inc., under
NASA contract NAS5-26555.}
 
\author{Marc Sarzi\altaffilmark{2,3}, Hans-Walter Rix\altaffilmark{3}, 
Joseph C. Shields\altaffilmark{4}, Greg Rudnick\altaffilmark{5,3},\\
Luis C. Ho\altaffilmark{6}, Daniel H. McIntosh\altaffilmark{5}, 
Alexei V. Filippenko\altaffilmark{7}, Wallace L. W. Sargent\altaffilmark{8}} 

\altaffiltext{2}{Dipartimento di Astronomia, Universit\`a di Padova,
Vicolo dell'Osservatorio 5, I-35122 Padova, Italy}
\altaffiltext{3}{Max-Planck-Institut f{\"u}r Astronomie,
K{\"o}nigstuhl 17, Heidelberg, D-69117, Germany; rix@mpia-hd.mpg.de}
\altaffiltext{4}{Physics \& Astronomy Department, Ohio University,
Athens, OH 45701; shields@phy.ohiou.edu}
\altaffiltext{5}{Steward Observatory, University of Arizona, Tucson,
AZ 85721; dmac, grudnick@as.arizona.edu}
\altaffiltext{6}{The Observatories of the Carnegie Institution of
Washington, 813 Santa Barbara St., Pasadena, CA 91101-1292;
lho@ociw.edu}
\altaffiltext{7}{Astronomy Department, University of California,
Berkeley, CA 94720-3411; alex@astro.berkeley.edu}
\altaffiltext{8}{Palomar Observatory, Caltech 105-24, Pasadena, CA
91125; wws@astro.caltech.edu}

\begin{abstract} 
We present spatially extended gas kinematics at parsec-scale
resolution for the nuclear regions of four nearby disk galaxies, and
model them as rotation of a gas disk in the joint potential of the
stellar bulge and a putative central black hole.
The targets were selected from a larger set of long-slit spectra
obtained with the {\sl Hubble Space Telescope} as part of the Survey
of Nearby Nuclei with STIS (SUNNS).
They represents the 4 galaxies (of 24 ) that display symmetric gas
velocity curves consistent with a rotating disk.
We derive the stellar mass distribution from the STIS acquisition
images adopting the stellar mass-to-light ratio normalized so as to
match ground-based velocity dispersion measurements over a large
aperture.
Subsequently, we constrain the mass of a putative black hole by
matching the gas rotation curve, following two distinct approaches.
In the most general case we explore all the possible disk
orientations, alternatively we constrain the gas disk orientation from
the dust-lane morphology at similar radii.
In the latter case the kinematic data indicate the presence of a
central black hole for three of the four objects, with masses of $10^7
- 10^8$M$_\odot$, representing up to $\sim 0.025$\% of the host bulge
mass.
For one object (NGC~2787) the kinematic data alone provide clear
evidence for the presence of a central black hole even without
external constraints on the disk orientation.
These results illustrate directly the need to determine black-hole
masses by differing methods for a large number of objects, demonstrate
that the variance in black hole/bulge mass is much larger than
previously claimed, and reinforce the recent finding that the
black-hole mass is tightly correlated with the bulge stellar velocity
dispersion $\sigma$.
\end{abstract} 
  
 
%
%
\section{Introduction} 
\label{sec:SMBH_intro} 
 
Over the last few years evidence has mounted that supermassive black
holes (SMBH) are nearly ubiquitous in galactic centers.
This finding has resulted from high-resolution observations of various
kinematical tracers of the central gravitational potential, such as
stars, ionized gas, or water masers (see, e.g., Ho 1999 for a review).
Additional support for this picture is provided by spectroscopic
surveys demonstrating that weak active galactic nuclei (AGN) activity
is common in nearby galaxies, especially among early-type systems
(E-Sb; Ho, Filippenko, \& Sargent 1997a and references therein).  An
important fraction of these nuclei further exhibit broad-line emission
and/or compact X-ray or radio sources that are the hallmarks of quasar
activity, presumably powered by accretion onto a black hole (e.g., Ho,
et al. 1997b; Terashima, Ho, \& Ptak 2000; Nagar et al. 2000).
 
If SMBHs are indeed an integral part of galaxies and their formation,
a natural question is whether the black-hole mass $M_\bullet$ is
related to larger properties of the host galaxy.  A correlation
between $M_\bullet$ and stellar mass of the spheroidal component
$M_{bulge}$ was suggested by Kormendy (1993), and later quantified by
Kormendy \& Richstone (1995), Magorrian et al. (1998), and Ho (1999).
As investigated recently by Kauffmann \& Haehnelt (2000), such a
correlation might arise naturally in hierarchical scenarios for galaxy
formation, and is expected if local SMBHs are the dormant relics of
past quasar activity.  Evidence of a close connection between galaxy
structure and SMBHs has been bolstered by recent reports of a
correlation between $M_\bullet$ and the velocity dispersion of the
surrounding stellar bulge (Ferrarese \& Merritt 2000, hereafter FM00;
Gebhardt et al. 2000a, hereafter G00).
 
The largest compilation of SMBH mass estimates obtained with a single
technique, namely axisymmetric two-integral models for the central
stellar kinematics, is that of Magorrian et al. (1998).  For most of
the 32 early-type galaxies of their sample, they found that {\sl
Hubble Space Telescope} ({\sl HST}) photometric and ground-based
kinematical data were consistent with the presence of a central
supermassive black hole.
Based on their data they claim that $M_\bullet \simeq 0.006
$M$_{bulge}$ with a scatter of only a factor of $\sim 3$.  While the
restrictive orbital assumptions in their modeling can lead to an
overestimate of the black-hole mass (Magorrian et al. 1998), their
high detection rate nonetheless underscores the apparent prevalence of
SMBHs and their close connection to the surrounding stellar bulge.
 
Despite this recent progress, the demography of SMBHs is far from
complete.
Most of the reported SMBHs have been detected in bulge-dominated
systems, and it is not clear if the $M_\bullet/M_{bulge}$ relation
holds also in the case of disk-dominated galaxies.  This issue is
important in light of theories for bulge formation via secular
evolution of disk structures (e.g., Norman, Sellwood, \& Hasan 1996).
In addition, it is still not fully understood whether the apparent
lack of low-mass central black holes in very massive galaxies is due
to selection effects, in which case the correlated distribution of
points in the $M_\bullet/M_{bulge}$ plane actually represents only an
upper envelope.
In Seyfert nuclei and quasars, measurements of $M_\bullet$ derived
from reverberation mapping experiments provide intriguing evidence
that the average $M_\bullet/M_{bulge}$ ratio in active galaxies is
systematically lower than in normal ones (Ho 1999; Wandel 1999; Kaspi
et al. 2000, Gebhardt et al. 2000b), even though this discrepancy may
disappear in the $M_\bullet/\sigma$ ratio (Gebhardt et al. 2000b).
Independent determinations of $M_\bullet$ in active galaxies are
desirable for confirmation of these results.  Black-hole mass
estimates and the corresponding Eddington luminosity $L_{Edd}$ are
also of fundamental importance for testing theories of accretion
physics in AGNs, since models typically predict a strong dependence of
accretion disk structure on $\dot M/M_\bullet$, or alternatively
$L/L_{Edd}$ (e.g., Narayan, Mahadevan, \& Quataert 1998).
 
In order to address these and other issues we initiated a Survey of
Nearby Nuclei with STIS (SUNNS) on {\sl HST}, with kinematic
information obtained from measurements of nebular emission lines.
Because of its collisional nature, gas promises a conceptually easier
way to trace the gravitational potential of a galaxy nucleus than
stars, because nearly circular orbits can be assumed; at the same
time, however, gas is more susceptible to non-gravitational forces and
may not be in equilibrium.  
Use of emission-line tracers to constrain a central mass concentration
is thus limited to cases where strong evidence exists that the gas
dynamics are dominated by gravity.
 
The paper is organized as follows. In \S 1 we present the
spectroscopic and photometric STIS observations, and in \S 2 we
describe our modeling of the ionized gas kinematics.  We present our
results in \S 3 and draw our conclusions in \S 4.
 
%
%
\section{Observations and data reduction} 
\label{sec:SMBH_obs} 

In carrying out SUNNS, long-slit spectra were obtained with STIS for
the central regions of 24 nearby, weakly active galaxy nuclei.  Full
details of the observations and data calibration are reported by Rix
et al. (2001 {\it in prep}). The observations were acquired in 1998
and 1999, with the $0\farcs2 \times 52\arcsec$ slit placed across each
nucleus along an operationally determined position angle (P.A.), which
is effectively random with respect to the projected major axis of any
galaxy subsystem.  After initial 20-s acquisition exposures with the
Long Pass filter (roughly equivalent to $R$), from which we derive
surface photometry of the nuclei, two spectral exposures totaling
$\sim 30$ minutes and three exposures totaling $\sim 45$ minutes were
obtained with the G430L and G750M gratings, respectively.  The
resulting spectra span 3300 -- 5700 \AA\ and 6300 -- 6850 \AA\ with
spectral resolution for extended sources of 10.9
\AA\ and 2.2 \AA , respectively.  The telescope was offset by 0\farcs05 
($\approx$ 1 pixel) along the slit direction between repeated
exposures to aid in the removal of hot pixels.

The two-dimensional (2-D) spectra were bias- and dark-subtracted,
flat-fielded, aligned, and combined to single blue and red spectra,
cleaned of remaining cosmic rays and hot pixels, and corrected for
geometrical distortion.  The data were wavelength- and flux-calibrated
with standard STSDAS procedures.  The exact reduction steps differed
slightly from image to image, in particular the image combining,
dithering, and hot pixel removal.  A detailed discussion of the
reduction steps and the 2-D spectra will appear in Rix et al. (2001
{\it in prep}).  For the kinematic analysis reported here, we used
only the red, high-resolution spectra.  We determined the radial
velocity of the ionized gas as a function of position along the slit
from Doppler shifts of the H$\alpha$ and [\ion{N}{2}]
$\lambda\lambda$6548, 6583 emission lines.  Measurement of these
features was carried out using the program SPECFIT as implemented in
IRAF (Kriss 1994), assuming a common velocity frame and Gaussian
profile for the three lines, and a [\ion{N}{2}] doublet ratio (1:3)
dictated by atomic parameters.  SPECFIT employs $\chi^2$-minimization
methods that provide error estimates for the measured velocities.

\begin{figure} 
\plotone{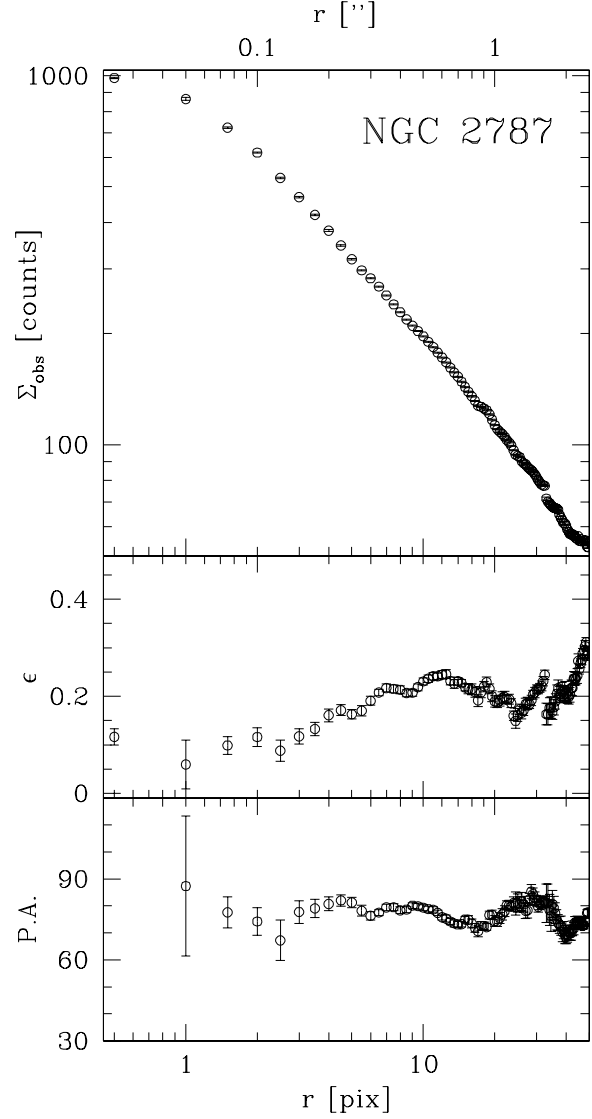}
\caption{Radial profiles of the surface brightness, ellipticity and position angle in
NGC~2787, NGC~4203, NGC~4459 and NGC~4596 obtained from the STIS broad-band acquisition images.}
\label{fig:SMBH_in_bulges_fig1}  
\end{figure}

Of the target galaxies, 12 displayed spatially resolved nebular
emission at a level that was detectable in our data. For these
sources, many of the measured gas velocity profiles showed distinct
asymmetries and other features that appear inconsistent with
steady-state orbital motion.  
We consequently selected for further study only those objects for
which the gas velocity curves were extended, single valued, and nearly
reflection-symmetric about the center of the galaxy; six objects met
these criteria.
We subsequently eliminated two of the remaining objects (NGC~3351 and
NGC~4548) from the present study since surface photometry indicates
that their central regions exhibit large deviations from sphericity or
strong isophotes twisting.
We defer discussion of these sources to a later analysis. The
remaining four objects that we present here are NGC~2787, NGC~4203,
NGC~4459, and NGC~4596.
Global parameters for these galaxies are given in Table
\ref{tab:SMBH_in_bulges_basic_parameters}, while their surface 
brightness, ellipticity, and position angle profiles are shown in
Fig.\ref{fig:SMBH_in_bulges_fig1}.
These radial profiles were obtained with the task ELLIPSE in IRAF,
after masking the most prominent dust features (e.g. NGC~2787).  We
adopted ellipses with linearly spaced semi-major axes, with a minimum
radius of 0.5 pixel.

\placetable{tab:SMBH_in_bulges_basic_parameters}
\placefigure{fig:SMBH_in_bulges_fig1}

%
%
\section{The modeling of the ionized gas kinematics} 
\label{sec:SMBH_model} 
 
\noindent 
We used the data to derive constraints on the presence and mass of a
SMBH by modeling the gas kinematics in the combined gravitational
potential of the stars and of a central dark mass.  The projected
light distribution in the centers of these galaxies is relatively
round (axis ratio $b/a \ge 0.7$), and we consequently assumed
spherical symmetry for the stellar mass distribution.  This
approximation is sufficient to roughly estimate the {\it stellar}
mass-to-light ratio within $ \la 3\arcsec$.  We further assumed that
the circumnuclear ionized gas is moving on planar, closed and hence
nearly circular orbits in the total gravitational potential.
The spatial gas distribution was taken to be a disk, whose orientation
is specified by its inclination $ i $ with respect to the sky plane,
and by the angle $ \phi $ between the directions of its projected
kinematic major axis and the slit position.
 
To derive the stellar portion of the gravitational potential, we
proceeded in several steps:
 
\subsection{Stellar Mass Profile} 
 
We derived the deprojected stellar luminosity density $\nu(r)$ from
the observed surface brightness profile $\Sigma_{\rm obs}(R)$ in the
STIS acquisition image.
The intrinsic surface brightness profile $\Sigma(R)$ was modeled as a
sum of Gaussian components (Monnet, Bacon, \& Emsellen 1992), which
were convolved with the {\sl HST} point-spread function (PSF).
The PSF itself was also represented as a sum of Gaussians, and was
derived from stars present in the STIS acquisition images.  This
yielded a PSF very similar to the synthetic ones obtained using the
Tiny Tim package (Krist \& Hook 1999) and with stellar spectral energy
distributions.
The convolved model was then compared with $\Sigma_{\rm obs}(R)$ in
order to determine the optimal scaling coefficients for the Gaussian
components.
In practice, we constrained the Gaussian width coefficients,
$\sigma_i$, to be a set of logarithmically spaced values, thus
simplifying the multi-Gaussian decomposition into a general
non-negative linear least-square problem for the corresponding
Gaussian amplitudes, $\alpha_i$.
The resulting deconvolved $\Sigma(R)$ is then readily deprojected to
obtain $\nu(r)$. For a radially constant mass-to-light ratio
$\Upsilon$, a multi-Gaussian description also results for the stellar
mass density $\rho (r)=\Upsilon\!\cdot\!\nu (r)$, whose contribution
$\Phi_{\star}(r)$ to the total potential $\Phi (r)$ can be computed
conveniently in terms of error functions.

In applying this method to NGC~2787, NGC~4203, and NGC~4596, we found
a point-like, presumably non-thermal light component in addition to
the extended stellar light.
We note that the absence of such a component in NGC~4459 is consistent
with its comparatively weak indications of AGN activity, reflected in
its spectroscopic classification as an \ion{H}{2}/LINER transition
object.
Central point sources manifest themselves by abrupt changes in the
surface brightness profile slope in the innermost 1-2 pixels, and are
well fit by PSF-convolved point sources.
We consequently removed this component from the measured $\Sigma_{\rm
obs}(R)$ in order to measure only stellar light.
As an example, in Figure \ref{fig:SMBH_in_bulges_fig2} we show the
multi-Gaussian fit to $\Sigma_{\rm obs}(R)$ for NGC~4203, along with
the recovered luminosity density profile $\nu(r)$ and the shape of the
corresponding circular velocity curve $V_c(r)$.
  
\placefigure{fig:SMBH_in_bulges_fig2}
 
\subsection{Stellar Mass-to-Light Ratio} 
 
Values of $\Upsilon$ can be constrained dynamically using measurements
of the stellar velocity dispersion $\sigma(r)$ averaged over a size
scale sufficiently large that a plausible SMBH will not dominate the
potential.  
We used ground-based measurements through apertures of several square
arcseconds for this purpose (Table
\ref{tab:SMBH_in_bulges_basic_parameters}).  We assumed a constant
$\Upsilon$ for each galaxy and solved the Jeans equation in the
spherical isotropic case (Binney \& Tremaine (1987)) for the velocity
dispersion profile $\sigma(r)$.
We then integrated along the line of sight to obtain the projected
$\sigma_p(R)$, and computed a seeing-convolved,
surface-brightness-weighted mean $<\!\sigma_p(R)\!>$ within the given
aperture.
Finally, we adjusted $\Upsilon$ until $<\!\sigma_p(R)\!>$ matched the
observed $\sigma$.

As a refining step, we included the effect of a central black hole in
the on $<\!\sigma_p(R)\!>$ producing a $\Upsilon(M_\bullet)$ curve,
which is monotonically decreasing with increasing black-hole masses,
from the $\Upsilon(0)$ maximum value without any central mass
concentration down to the limiting zero value corresponding to some
maximum allowed black-hole mass (typically $\sim 10^9$ M$_\odot$).

All the model steps, including the multi-Gaussian fit of $\Sigma(R)$,
recovery of $\nu(r)$, calculation of $V_c(r)$, and reconstruction of
$\sigma(r)$ and $\sigma_p(R)$, were tested and verified using
Hernquist (1990) analytical models for all these functions.
 
\subsection{Gas Velocity Field} 
 
The prediction of the observed gas velocity field for our galaxies
depends on the orientation of the gas disk in each case.
We cannot expect that the parsec-scale central gas disk is coplanar
with the vastly larger stellar galaxy disk; indeed, there are examples
were this is clearly not the case (e.g., NGC~3227; Schinnerer et
al. 2000).
We therefore modeled the gas kinematics with the disk inclination
angles $i$ and $\phi$ for the gas initially unconstrained.

Inspection of acquisition and archival WFPC2 images for our sources
reveals clear signatures of compact dust lanes.
As this dust and the ionized gas are most likely part of the same
central disk (e.g., Pogge et al. 2000, Verdoes Kleijn et al. 1999),
the dust-lane morphology provides a potential means of deriving the
disk orientation.
We obtained corresponding estimates of $i$ and $\phi$ for the gas disk
by defining ellipses consistent with the dust-lane morphology, and
assuming that deviations from circularity result from inclination and
projection.
An objective ellipse-fitting algorithm for the dust features is
difficult to construct, since the strength of the dust lanes varies
spatially and is not continuous.
We therefore constructed ellipses by eye that are consistent with the
dust-lane morphology, with the outcome shown in Figure
\ref{fig:SMBH_in_bulges_fig3}.
These results were checked with absorption maps generated by
subtraction of a smooth starlight profile from the original images.
Estimates of ellipse parameters obtained independently by different
members of our team (MS and HWR) showed good agreement ($\Delta(b/a)
\la 0.1$ and $\Delta({\rm P.A.}) \la 10^\circ$), providing some
confidence in these fits.
Note, that all the dust rings seem quite highly inclined.  In this
inclination regime $\sin{i}$ is not very sensitive to errors in $i$.
We used the results to generate a second, constrained set of models
with the disk orientation specified by the dust morphology.

\placefigure{fig:SMBH_in_bulges_fig3}

For a given $i$ and $\phi$ and a given total potential $\Phi(r) =
\Phi_\bullet(r) + \Phi_*(r)$, the projected line-of-sight gas velocity 
field $V_p(x_p,y_p)$ is completely specified by our model.
To make observable predictions, this velocity field must be further
weighted by the spatial distribution of line flux, and convolved by
the {\sl HST} PSF.
Directly observed gas disks on parsec scales (e.g., M87, Harms et
al. 1994) show that the gas may be patchy.
Given our limited data, our best guess for the two-dimensional gas
emissivity distribution $f(x_p,y_p)$ is still the axisymmetric
extension of the deconvolved line flux profile along the slit.
The latter was obtained also through a multi-Gaussian deprojection.
The observable velocity field resulted then from convolving
$f(x_p,y_p)\cdot V_p(x_p,y_p)$ with the PSF, and sampling the result
along the slit. This procedure provides model points for direct
comparison with our STIS measurements.
 
%
%
\section{Modeling Results} 
\label{sec:SMBH_results} 

\subsection{Model--Data Comparison} 
In the unconstrained case, we considered all possible orientations, or
the full $(i,\phi)$ parameter space. For any given black-hole mass
$M_\bullet$ and its corresponding $\Upsilon(M_\bullet)$, we identified
the set of $(i,\phi)$ that best matches the observed kinematics.
The quality of this match was quantified by
$\chi^2_{free}(M_\bullet|\Upsilon(M_{\bullet}))$ values, each
optimized over $(i,\phi)$.

For fits with fixed disk orientation, we simply adopted the values
from the dust lane morphology and explored $\chi^2_{fix}(M_\bullet|
\Upsilon(M_\bullet))$ for a {\it fixed} $(i,\phi)$. 
By constructing $\chi^2_{free}$ and $\chi^2_{fix}$ curves we can check
more stringently the necessity for a central mass concentration,
eventually deriving the best $M_\bullet$ values for each case.
 
We considered a number of errors and modeling uncertainties and
assessed their importance for the $M_\bullet$ estimates through
simulations. This analysis considered uncertainties in (1) the derived
$\nu(r)$, particularly at small $r$, (2) the multi-Gaussian fit to the
surface brightness of the ionized gas, (3) the location of the
kinematical center of the observed rotation curves, and (4)
$\Upsilon(M_\bullet)$, including the effects of possible deviations
from our assumption of sphericity and isotropy for the stellar
distribution, and the variance among published values for the central
velocity dispersions.
Neglecting the intrinsic flattening can lead us underestimate the
mass-to-light ratio $\Upsilon$; from the tensor virial theorem the
fractional error on $\Upsilon$ can be evaluated to be less then half
of the intrinsic axis ratio $c/a$ (Kronawitter et al. 2000).  Assuming
for our four galaxies a typical apparent flattening in their central
regions of $b/a=0.85$ (Fig. \ref{fig:SMBH_in_bulges_fig1}) and an
inclination of 40$^{\circ}$
(Tab. \ref{tab:SMBH_in_bulges_basic_parameters}), their intrinsic axis
ratios would be around $c/a=0.6$, and the mass-to-light ratio would
need to be increased by $21\%$ (Kronawitter et al. 2000).  Note that
the published values for the central velocity dispersion of our sample
galaxies have a variance of $3\%-20\%$.

This last factor proved to have the largest impact on $M_\bullet$.
Therefore, we obtained two additional fitting sequences
$\chi^2(M_\bullet |\Upsilon(M_\bullet ))$, adopting stellar
mass-to-light ratios equal to 0.7 and 1.3 times the derived
$\Upsilon(M_\bullet )$, both in the unconstrained and constrained disk
orientation cases. In essence, we are hereby allowing that our simple
modeling has introduced 30\% {\it systematic} errors in the
$\Upsilon(M_\bullet )$ determination.
 
The rotation curves in our culled sample show small wiggles and
asymmetries in excess of their statistical errors, presumably due to
non-gravitational phenomena acting on the gas. In practice, this means
that we cannot expect to generate a model that is formally acceptable
in a $\chi^2$ sense, and the estimation of confidence limits is
correspondingly affected.  To allow for the additional velocity
structure that is not addressed by our model, we rescaled all $\chi^2$
values to achieve $\chi^2_{free}$ = $N_{DOF}\equiv N_{Data} - N_{fit}$
for the best unconstrained fit, where $N_{DOF}$, $N_{Data}$, and
$N_{Fit}$ are the numbers of degrees of freedom, data points, and fit
parameters, respectively.  Note that this procedure is conservative in
the sense of widening the confidence intervals.
Confidence limits on the range of $M_\bullet$ were then derived from a
$\Delta\chi^2$ test (e.g. Press, et al. 1986), for the variation of
one parameter.
 
\subsection{Fits with Unconstrained Disk Orientation}
\noindent 
The results of the unconstrained fits are shown in the top and middle
panels of Figures \ref{fig:SMBH_in_bulges_fig4}$a-d$,
and are summarized in Table \ref{tab:SMBH_in_bulges_modeling_results}.
The resulting uncertainties in $M_\bullet$ are large, and only one of
the objects (NGC~2787) exhibits evidence of a nonzero mass at more
than the $2\sigma$ level.
In all sample galaxies the gas rotation curve extends beyond the radii
where the black hole is likely to dominate; there the observed gas
motion should approach the one predicted from the stellar mass, which
was obtained from the stellar modeling.  This requirement provides a
constraint on the gas inclination, and precludes arbitrarily large
$M_\bullet$ with disks seen at large $i$. 
In this way, we can obtain robust upper limits on $M_\bullet$, which
prove interesting in some cases such as NGC~4203 ($M_\bullet < 5\times
10^6 \; $M$_\odot$ at $3\sigma$), which is discussed in detail by
Shields et al. (2000).  We note that the limiting value presented here
for this object is 20\% smaller than the value reported by Shields et
al. (2000); this difference is due to the use of $\Upsilon(M_\bullet)$
in the present analysis, rather than a mass-to-light ratio that
neglects the presence of the black hole.
 
\placefigure{fig:SMBH_in_bulges_fig4}

\subsection{Fits with Fixed Disk Orientation} 

The results of the fits with constrained disk inclination are shown in
the top and lower panels of Figures
\ref{fig:SMBH_in_bulges_fig4}$a-d$, and are reported in
Table \ref{tab:SMBH_in_bulges_modeling_results}.  We note that in all
cases except NGC~4203, the constrained fit is statistically comparable
(in a $\chi^2$ sense) to the unconstrained fit and shows evidence for
a central mass concentration at a $< 2\sigma$ level. 
The quality of the fit is thus not improved by relaxing the disk
orientation while maintaining $M_\bullet$ at the value dictated by the
minimum $\chi^2_{fix}$. This finding is indicated in Figures 2$a,c,d$
by the fact that the thick curves representing $\chi^2_{fix}$ coincide
with the thin curves where $\chi^2_{fix}$ attains a minimum. 
This result means that the unconstrained procedure leads to ($i,\phi$)
values very similar to the one derived from the dust patterns, providing
strong supporting evidence that this is indeed the correct choice of 
orientation.

Not surprisingly, Figure \ref{fig:SMBH_in_bulges_fig4}
shows that use of the fixed disk orientation constrains $M_\bullet$
much more tightly, with $M_\bullet \approx 7 \times 10^7$ M$_\odot$
for each of NGC~2787, NGC~4459, and NGC~4596, the individual values
and their uncertainties are listed in table
\ref{tab:SMBH_in_bulges_modeling_results}. 
These errors reflect the $1\sigma$ confidence regions for
$M_{\bullet}$ from the constrained fit after considering all the three
alternative values of $\Upsilon(M_\bullet)$.  For NGC~4459 Bertola et
al. (1998) derived from ground-based observations an upper-limit for
$M_\bullet$ of $10^9$ M$_\odot$ with an error of a factor of 3.
For NGC~4203 the best agreement was found with $M_\bullet = 5.2 \times
10^7$ M$_\odot$.  For this last object, the minimum $\chi^2_{fix}$ is
dramatically worse than $\chi^2_{free}$ at the same $M_\bullet$, and
depends sensitively on the assumed angle $\phi$; in our constrained
case, $\phi = 75^\circ$, representing a slit PA near the minor axis of
the disk.  None of the models for this source adequately reproduces
both the inner and outer parts of the rotation curve in detail.
The constrained fit appears consistent at large radii ($r \ga
0\farcs7$), but invariably shows substantial deviations in the central
region that produce the large $\chi^2_{fix}$. Since the outer disk in
this model appears consistent with the orientation selected by the
dust lanes, which reside at even somewhat larger radii, and taking
into account that the innermost part of the rotation curve is nicely
matched in the unconstrained case, we speculate that the rotation
curve is affected by a warp in the gas disk in this source.
Besides, gas kinematics in other galactic nuclei have been
successfully modeled by taking warped structures into account (e.g.,
NGC~3227, Shinnerer et al. 2000).  In light of these difficulties with
the constrained fit, we adopt here only an upper limit on $M_\bullet$
based on the unconstrained model, as reported also by Shields et
al. (2000).
 
\placetable{tab:SMBH_in_bulges_modeling_results}

%
%
\section{Discussion} 
\label{sec:SMBH_discus} 
 
The results of this study are important in several ways for the
understanding of SMBHs in galaxies.  For all four galaxies analyzed
here, we find qualitative and/or quantitative evidence for the
existence of SMBHs at their centers.  For three sources we obtained
well-determined values of $M_\bullet$; for the fourth object,
NGC~4203, the likely presence of a SMBH is signaled by very broad
($\pm $3000 km s$^{-1}$) H$\alpha$ emission (Shields et al. 2000).
An important point therefore is that {\sl all} of the sources examined
here apparently harbor a SMBH, reinforcing indications that such
objects are an integral part of most galaxies.  We emphasize that our
four objects were selected from the SUNNS sample on the basis of the
angular extent ($>0.5\arcsec$) and symmetry of their velocity curves,
not on the existence of a steep central velocity gradient or other
criteria indicative of a SMBH.
We note also that the majority of objects with kinematic measurements
of $M_\bullet$ to date are elliptical galaxies (FM00, G00, and
references therein), so that the present results for four S0 galaxies
represent a substantial increase in the census of SMBHs in bulges of
disk galaxies.  The findings described here are also important for
comparison with suggested trends relating $M_\bullet$ to galaxy bulge
properties.  For our four galaxies, we obtained bulge $B$-band
luminosities using $V$-band disk--bulge decompositions from Baggett,
Baggett, \& Anderson (1998), and $B-V_e$ colors within the galaxy
effective radius, with the results listed in Table
\ref{tab:SMBH_in_bulges_DBdecomposition}.
Bulge masses were estimated using the mean $B$-band $\Upsilon = 8.90$
($M/L$)$_\odot$ from van der Marel (1991), rescaled to $H_0 = 75$ km
s$^{-1}$ Mpc$^{-1}$.
For comparison with FM00 and G00, we also calculated stellar velocity
dispersions $\sigma_e$ and $\sigma_{e/8}$ within the bulge effective
radius $R_e$ and within $R_e/8$ respectively, using values of $\sigma$
from Table \ref{tab:SMBH_in_bulges_basic_parameters}, seeing-corrected
measurements of $R_e$ from Baggett et al. (1998), and the algorithm of
Jorgensen, Franx, \& Kjaegaard (1995).

\placetable{tab:SMBH_in_bulges_DBdecomposition}
 
Our results for $M_\bullet$ as a function of bulge properties are
shown in relation to previous work in Figure
\ref{fig:SMBH_in_bulges_fig5}.
Compared with the $M_{bulge} - M_\bullet$ correlation reported by
Magorrian et al. (1998), our constrained results yield black-hole
masses that are lower by a factor of $\sim 3$ or more at a given
$M_{bulge}$ while two out of four of our unconstrained 3$\sigma$ upper
limits equally fall below the $M_{bulge} - M_\bullet$ relation,
consistent with other suggestions that the two-integral estimates of
$M_\bullet$ by these authors are systematically too large (e.g., van
der Marel 1998).
When compared with published analyses employing more robust measures
of SMBH masses, our results show better agreement, as can be seen in
the diagram of $M_\bullet$ versus $L_{bulge}$ from Ho (1999), which
derived $M_\bullet/L_{bulge}$ ratio translate in a
$M_\bullet/M_{bulge}\simeq0.002$ quite similar to the value first
inferred by Kormendy \& Richstone (1995).
The constrained fits in particular are in good agreement with the
latter correlation, although the upper limit for NGC~4203 continues to
suggest a non-negligible degree of scatter in the $M_\bullet$ versus
$L_{bulge}$ relation.

\placefigure{fig:SMBH_in_bulges_fig5}
 
The relationship between $M_\bullet$ and $\sigma$ is of particular
interest in light of recent reports by FM00 and G00 of a very
significant correlation with little intrinsic scatter.  Estimates of
stellar velocity dispersion employed by these authors correspond to
apertures large enough to avoid the influence of a black hole on
$\sigma$ itself. 
Even if taken {\it per se}, our data show no correlation between
M$_{bulge}$ and any of the other quantities considered so far, it is
worth noticing that while going from the unconstrained to the
constrained procedure, the black-hole mass determinations move closer
to the M$_{\bullet}$ {\it vs.}
L$_{bulge}$, $\sigma_{e}$, and $\sigma_{e/8}$ relations.  Our results,
and the constrained fits in particular, are in good agreement with the
quantitative relations given in both papers.  These findings provide
added support for the strength of an $M_\bullet - \sigma$ trend, and
its persistence in bulges associated with disk galaxies.

As a final comment, we note that while the present work provides an
illustration of the power of gas kinematics for finding and
quantifying the masses of SMBHs, it also shows that a significant
investment of observing time and effort is necessary in order to yield
results.
In particular, the SUNNS survey demonstrates that only a modest
fraction ($\la 25$\%) of nearby galaxies displaying emission-line
nuclei in ground-based observations can be expected with {\sl HST} to
show spatially resolved nebular emission with regular kinematics
suitable for estimation of $M_\bullet$.  Determination of the gas
velocity field with high confidence requires 2-D mapping with multiple
slit positions (e.g., Barth et al. 2000), which is expensive in terms
of telescope time.
The alternative of employing long-slit observations at a single PA can
be employed with some effort to extract useful estimates of
$M_\bullet$, as demonstrated by the present work.  In particular the
analysis reported here further illustrates the power of using dust
morphology as an indicator of gas disk orientation, in the absence of
full 2-D velocity data.

\acknowledgments

This research was supported financially through NASA grant NAG 5-3556,
and by GO-07361-96A, awarded by STScI, which is operated by AURA, Inc., for
NASA under contract NAS5-26555.  M. Sarzi gratefully acknowledges N. Cretton
and W. Dehnen for valuable comments and suggestions on this work.

%
%

\clearpage
\begin{deluxetable}{lccclcllcl}
\footnotesize
\tablecaption{Selected Galaxies Basic Parameters \label{tab:SMBH_in_bulges_basic_parameters}}
\tablewidth{0pt}
\tablehead{
\colhead{Galaxy} & \colhead{Morph. Type} & \colhead{$B_{tot}$} & \colhead{i} & \colhead{P.A.} & 
\colhead{Nuclear Type} & \colhead{Dist} & \colhead{$\sigma$} & \colhead{STIS slit P.A.} & 
\colhead{Obs. Date} 
\nl
\colhead{(1)} & \colhead{(2)} & \colhead{(3)} & \colhead{(4)} & \colhead{(5)} & 
\colhead{(6)} & \colhead{(7)} & \colhead{(8)} & \colhead{(9)} & \colhead{(10)} 
} 
\startdata
NGC~2787 & SB(r)0+   & 11.82 & 50 & 117 & L1.9 & 13.0 & 210 &  33.2  &  1998 Dec 5  \nl  
NGC~4203 & SAB0-:    & 11.80 & 21 &  10 & L1.9 &  9.7 & 124 &  105.3 &  1999 Apr 18 \nl 
NGC~4459 & SA(r)0+   & 11.32 & 41 & 110 & T2:  & 16.8 & 189 &  92.9  &  1999 Apr 23 \nl
NGC~4596 & SB(r)0+   & 11.35 & 42 & 135 & L2:: & 16.8 & 154 &  70.3  &  1998 Dec 20 \nl
\enddata
\tablecomments{Cols. (2), (3) and (4): morphological type, total apparent magnitude 
$B_{tot}$, inclination and major axis P.A. from de Vaucouleurs et al. (1991). 
Col. (5): nuclear emission type from Ho et al. (1997a): H = H$_{\rm II}$ nucleus, 
S = Seyfert nucleus, L = LINER, and T = transition object. The number attached to 
the class letter designates the type, while the ``:'' and ``::'' indicate an uncertain 
or highly uncertain classification, respectively. 
Col. (6): distance from Tully (1988) in Mpc. Col. (7): Ground-based central stellar velocity 
dispersion $\sigma$ in $\rm km\;s^{-1}$, taken from Dalle Ore et al. (1991) for the 
first three sources and from Kent (1990) for NGC~4596; these authors obtained measurements 
through apertures of size $1\farcs5\times 4\farcs0$ and $5\farcs4\times 1\farcs5$, 
respectively.
Col. (8): P.A. of the STIS aperture. Col. (9): UT observation date.}
\end{deluxetable}

\clearpage
\begin{deluxetable}{llllll}
\footnotesize
\tablecaption{Model Results\label{tab:SMBH_in_bulges_modeling_results}}
\tablewidth{0pt}
\tablehead{
\colhead{Galaxy} & \colhead{$M_{\bullet,free}$} &  \colhead{$M_{\bullet,fix}$}  & 
\colhead{$i$} & \colhead{$\phi$}  & \colhead{M$_{sph}$}
\nl
\colhead{(1)} & \colhead{(2)} & \colhead{(3)} & \colhead{(4)} & \colhead{(5)} & 
\colhead{(6)}
}  
\startdata
NGC~2787 & 18.5$^{+50.8}_{-14.2}$ & 7.1$^{+0.7}_{-0.9}$ & 49 & 14 & 2.9 \nl  
NGC~4203 & $<0.5$                 & 5.2                 & 49 & 75 & 1.5 \nl 
NGC~4459 & 1.8$^{+19.2}_{-1.8}$   & 7.3$^{+1.4}_{-1.4}$ & 47 & 8  & 6.4 \nl
NGC~4596 & 0.8$^{+49.3}_{-0.8}$   & 7.8$^{+3.8}_{-3.3}$ & 40 & 41 & 8.6 \nl
\enddata
\tablecomments{ Cols. (2) and (3): $M_\bullet$ estimates or upper
limits in units of $10^7$M$_{\odot}$, along with $3\sigma$ and
$1\sigma$ confidence limits, for the unconstrained and constrained
models, respectively. The only exception is the upper limit for the
black-hole mass of NGC~4203, which was derived according to the
$3\sigma$ confidence limit.  
Cols. (4) and (5): angles describing gas disk orientation, in degrees,
for the constrained models.  Col. (6): bulge mass in units of
$10^{10}$M$_{\odot}$ derived from the luminosities in Table
\ref{tab:SMBH_in_bulges_DBdecomposition} and the mean $\Upsilon$
derived by van der Marel (1991).}
\end{deluxetable}

\clearpage
\clearpage
\begin{deluxetable}{ccccccc}
\footnotesize
\tablecaption{Bulge Properties of Selected Galaxies\label{tab:SMBH_in_bulges_DBdecomposition}}
\tablewidth{0pt}
\tablehead{
\colhead{Galaxy} & \colhead{$V_{tot}$} & \colhead{$V_{bulge}$} & \colhead{$B-V_e$} & 
\colhead{$L_{bulge}$ ($10^9L_{\odot}$)} & \colhead{$\sigma_e$} & \colhead{$\sigma_{e/8}$}
\nl
\colhead{(1)} & \colhead{(2)} & \colhead{(3)} & \colhead{(4)} & \colhead{(5)} & \colhead{(6)} 
& \colhead{(7)}
} 
\startdata
NGC~2787 & 10.76 &  11.17 & 1.10 & 3.25 & 184.9 & 206.5 \nl  
NGC~4203 & 10.86 &  11.33 & 0.98 & 1.74 & 109.9 & 122.7 \nl 
NGC~4459 & 10.37 &  10.99 & 0.98 & 7.16 & 166.6 & 186.1 \nl
NGC~4596 & 10.41 &  10.69 & 0.96 & 9.61 & 136.3 & 152.3 \nl
\enddata
\tablecomments{Col. (2): galaxy total V-band magnitude from de Vaucouleurs et al. (1991).
Col. (3): {\sl V}-band magnitude for galaxy the bulge light, obtained by subtracting 
from the total galaxy light the disk contribution inferred from Baggett et al. (1998).
Col. (4): galaxy $B-V_e$ effective colors from de Vaucouleurs et al. (1991).
Col. (5): bulge $B$-band luminosities. 
Cols. (6) and (7): stellar velocity dispersion $\sigma_{e}$ and $\sigma_{e/8}$ in 
km s$^{-1}$, within a circular aperture of radius R$_{e}$ or R$_{e/8}$ respectively.}
\end{deluxetable}

%
%
\begin{figure} 
\plotone{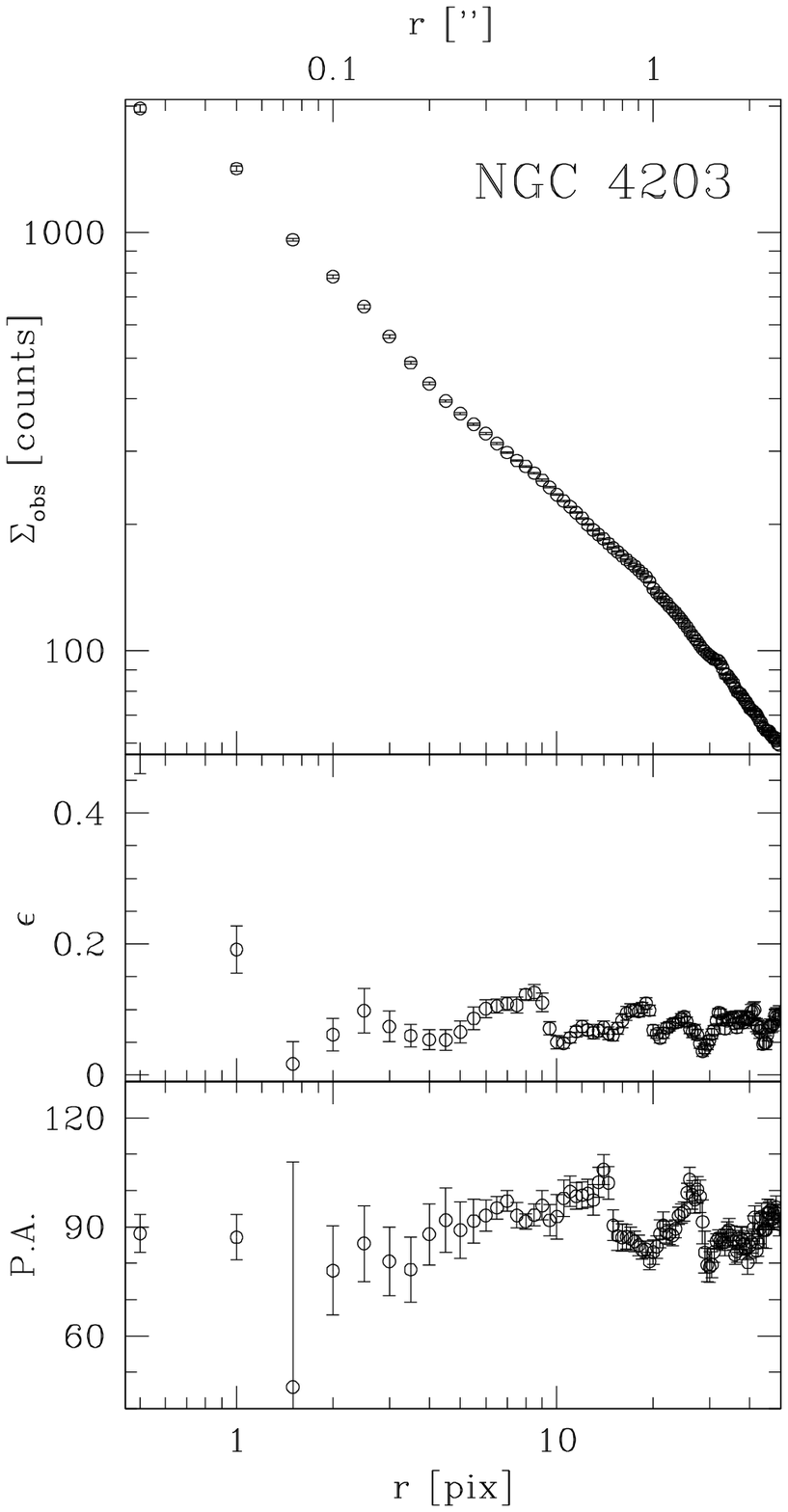}
\end{figure}  
\begin{figure} 
\plotone{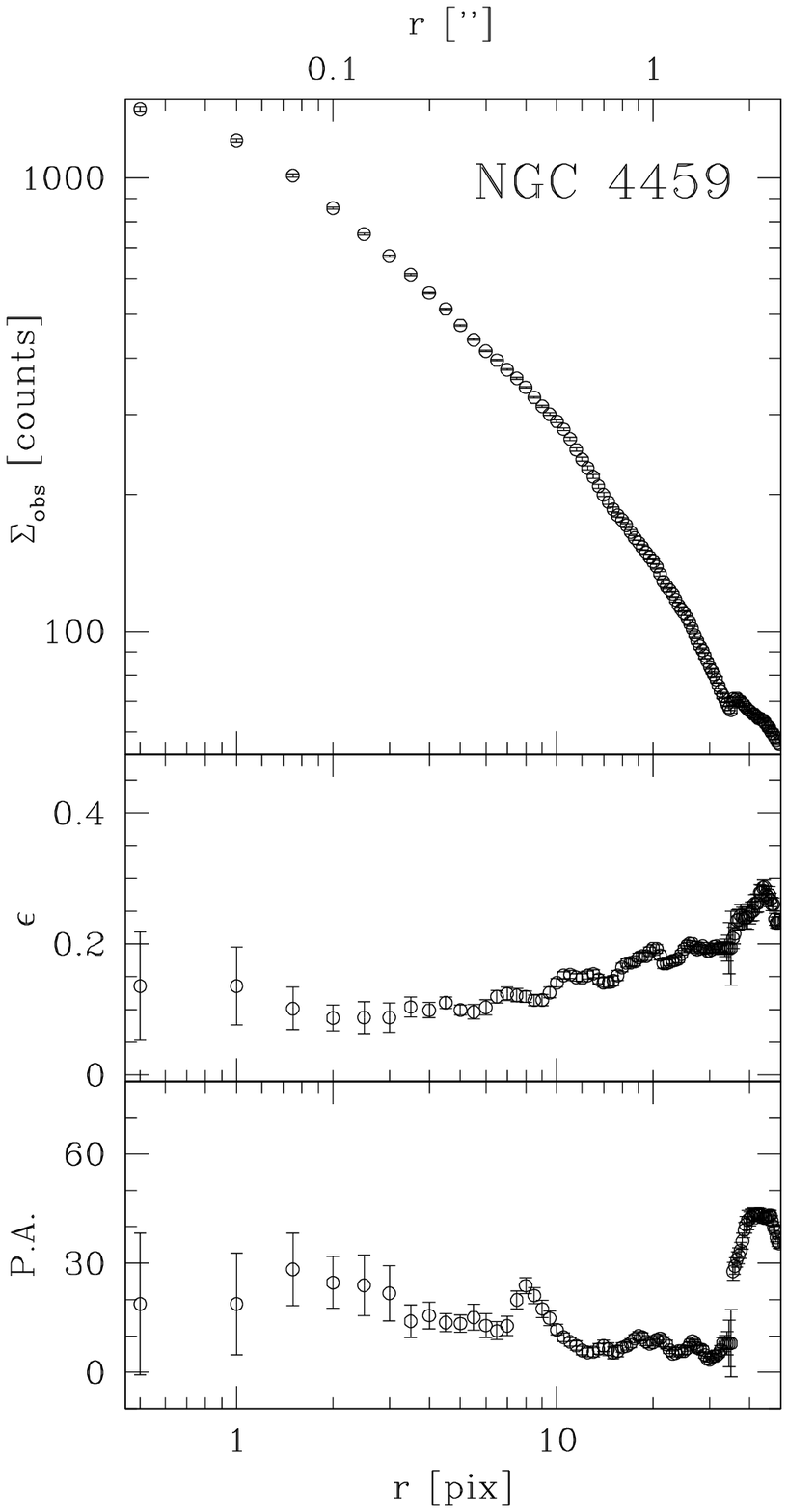}
\end{figure}  
\begin{figure} 
\plotone{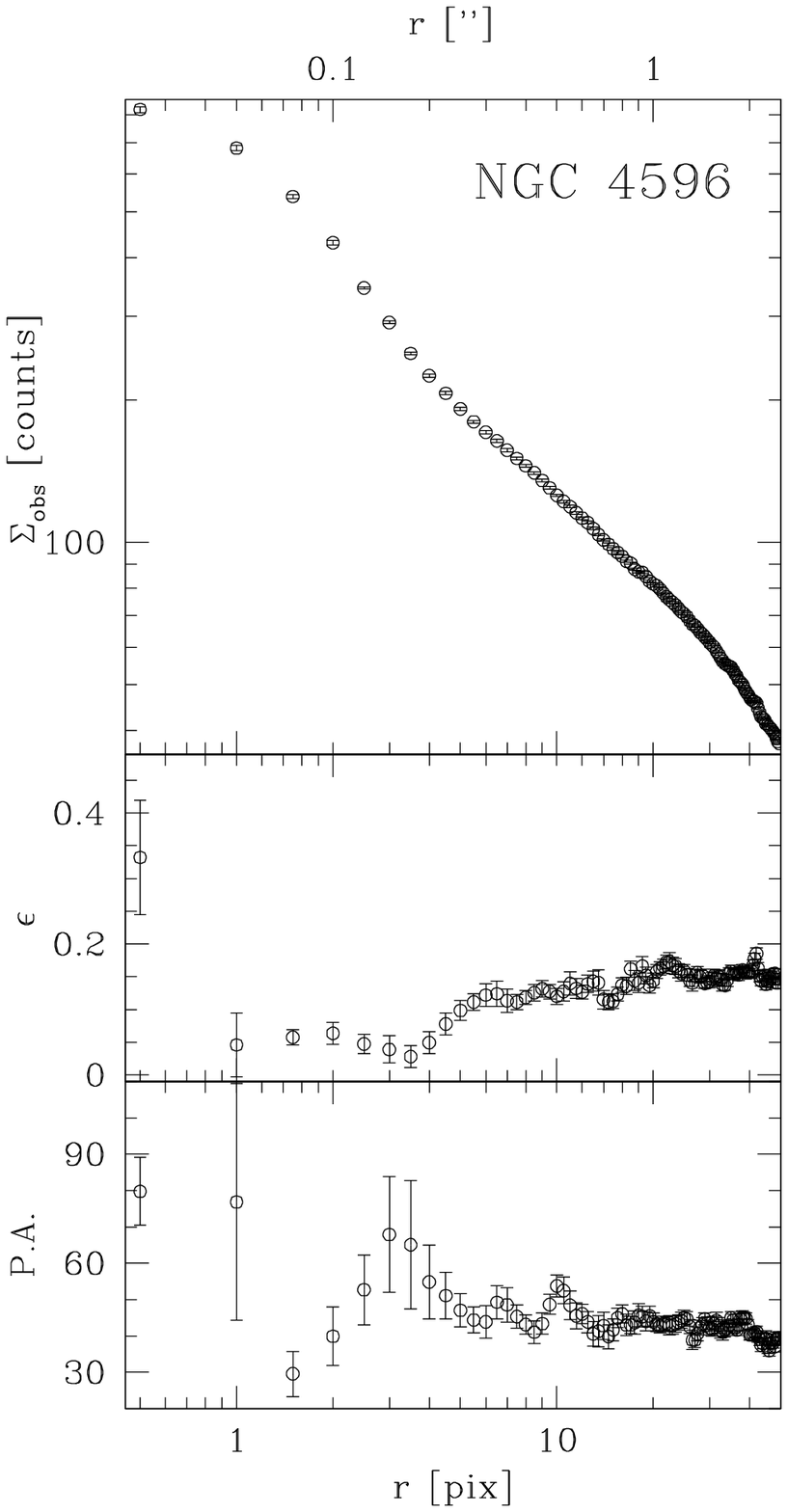}
\end{figure}  

\begin{figure} 
\plotone{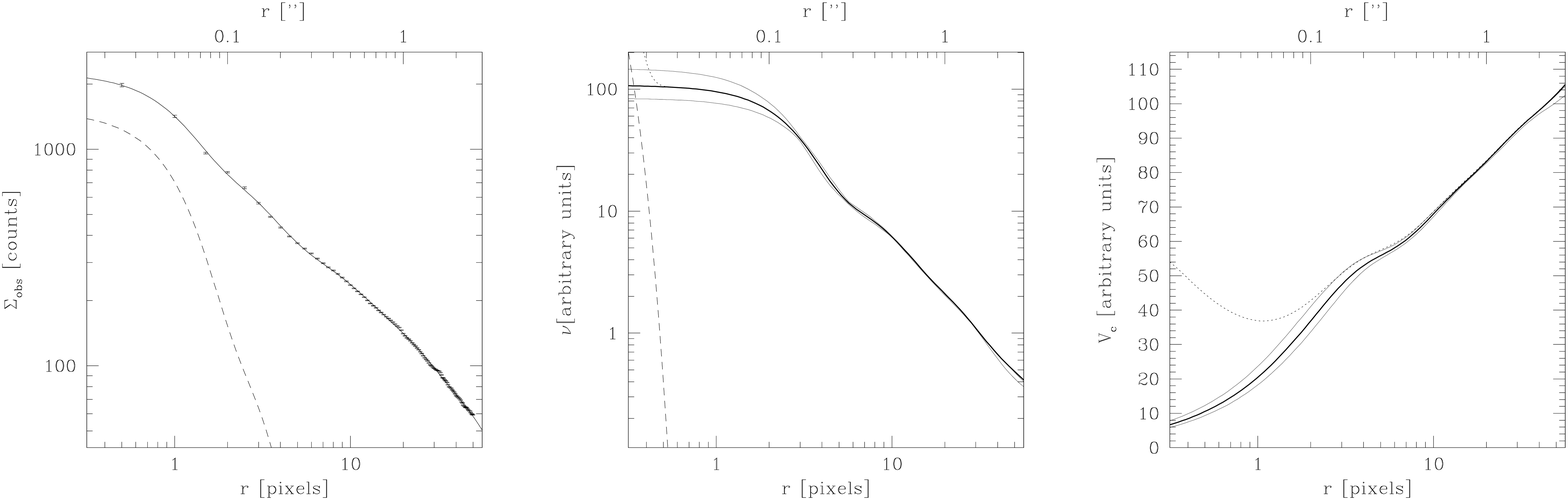}
\caption{Deprojection steps for the stellar mass profile of
NGC~4203: {\sl Left:} Multi-Gaussian fit to the observed surface
brightness $\Sigma_{\rm obs}(R)$ of NGC~4203 (solid line); the
inferred point-source contribution to the profile is shown by the
dashed line.  
{\sl Center:} Recovered intrinsic stellar luminosity profile $\nu(r)$
(thick solid line) after removal of the point source, and $3\sigma$
confidence limits on $\nu(r)$ (thin solid lines) obtained deprojecting
different Monte-Carlo realizations of $\Sigma_{\rm obs}(R)$.
Inclusion of this compact source (dashed curve) reproducing the
observations in their entirety would produce a narrow spike at the
center of $\nu(r)$ (dotted line).
{\sl Right:} Circular velocity profile $V_c(r)$ (thick solid line)
that results if the mass density is proportional to $\nu(r)$, along
with $3\sigma$ confidence curves (thin solid line).  The dotted line
shows the effect of assigning mass density proportional to the
luminosity density profile that includes the central compact feature.}
\label{fig:SMBH_in_bulges_fig2}  
\end{figure}  

\begin{figure} 
\plotone{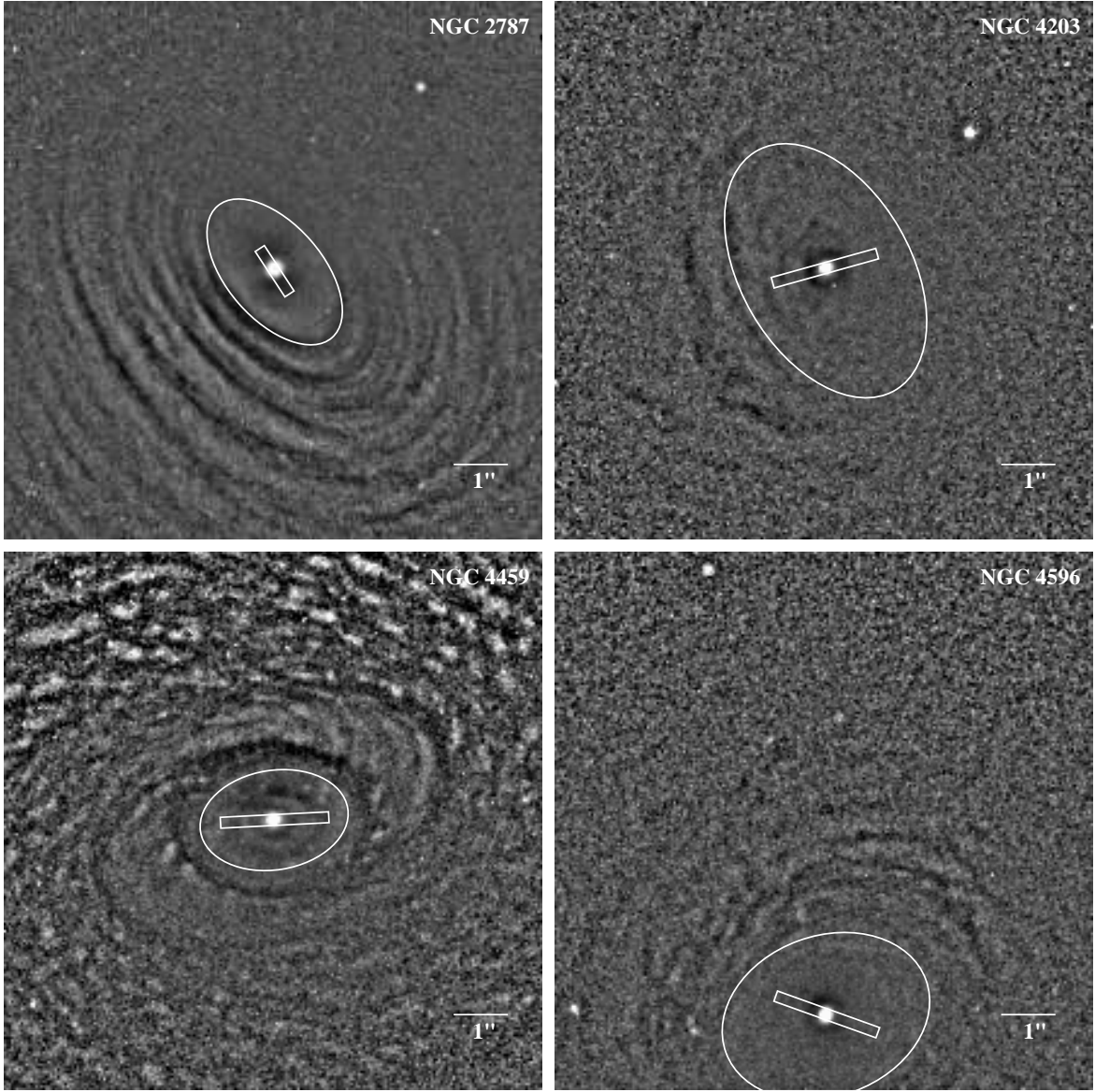}
\caption{$10\arcsec\times 10\arcsec$ unsharp-masked WFPC2 images for
the central regions of our modeled galaxies. North is up, east is
left. Superimposed on each frame is the ellipse showing the assumed
gaseous disk orientation and the STIS slit aperture, where the
depicted length corresponds to the maximum extent of our kinematic
measurements.  All images were derived from frames obtained in the
F555W passband, except for NGC~4596, for which the F606W filter was
used.}
\label{fig:SMBH_in_bulges_fig3}  
\end{figure}  
 
\begin{figure} 
\plotone{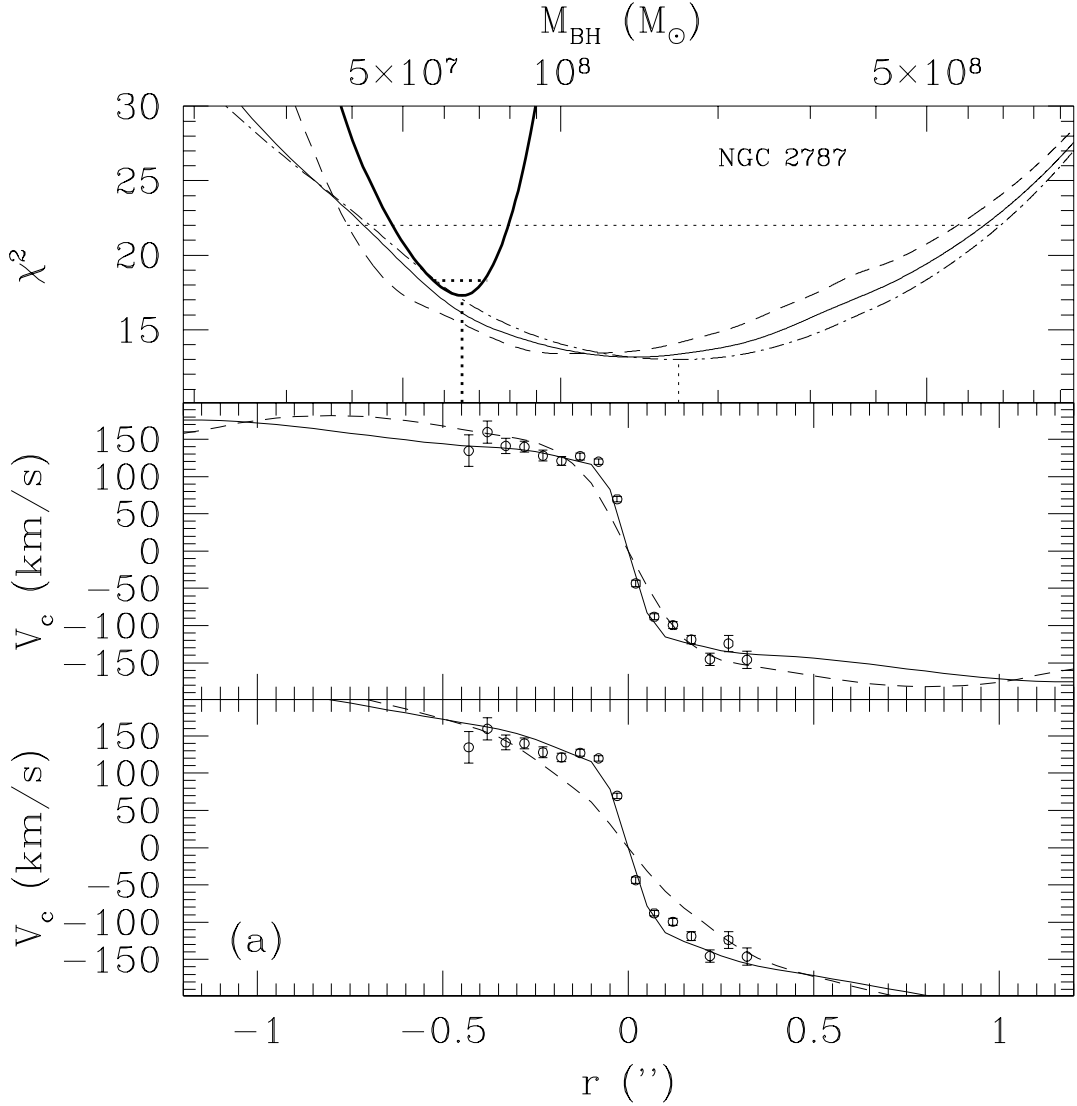}
\caption{Observed rotation curves along with the modeling results in 
both the unconstrained and constrained cases.
{\it Upper panels:} Rescaled $\chi^2_{free}$ (thin lines) and $\chi^2_{fix}$
(thick lines) as a function of $M_\bullet$.  The effect of adopting
alternative values of $\Upsilon(M_\bullet)$ rescaled by factors of
0.7, 1, and 1.3 are shown in the unconstrained case by the thin
dashed, solid, and dotted-dashed lines, respectively.
The thin dotted lines indicate the formally best estimate of $M_{\bullet}$
in the unconstrained case vertically, and the $3\sigma$ (upper) limits horizontally.
The thick dotted lines mark the estimate for $M_{\bullet}$ in the constrained 
case for the best $\Upsilon(M_\bullet)$ fitting sequence (vertically), along 
with the corresponding variances (horizontally).
The vertical and horizontal dotted lines indicate best values in the two cases for
$M_\bullet$ and corresponding $1\sigma$ and $3\sigma$ confidence limits,
for the constrained and unconstrained case, respectively. For NGC~4203 the 
$\chi^2_{fix}$ curve is off-scale in the vertical direction.
{\it Middle panels:} velocity as a function of position along the
slit, for the observed values (points), the unconstrained model
with no black hole (dashed lines), and the best-fit unconstrained
model including a black hole (NGC~2787, solid lines) or $M_\bullet$ set at the 
$3\sigma$ upper limit (dotted lines). 
{\it Lower panels:} velocity as a function of position along
the slit, for the observed values (points), the constrained model
with no black hole (dashed lines), and the best-fit constrained
model including a black hole (solid lines).}
\label{fig:SMBH_in_bulges_fig4}  
\end{figure}
\begin{figure} 
\plotone{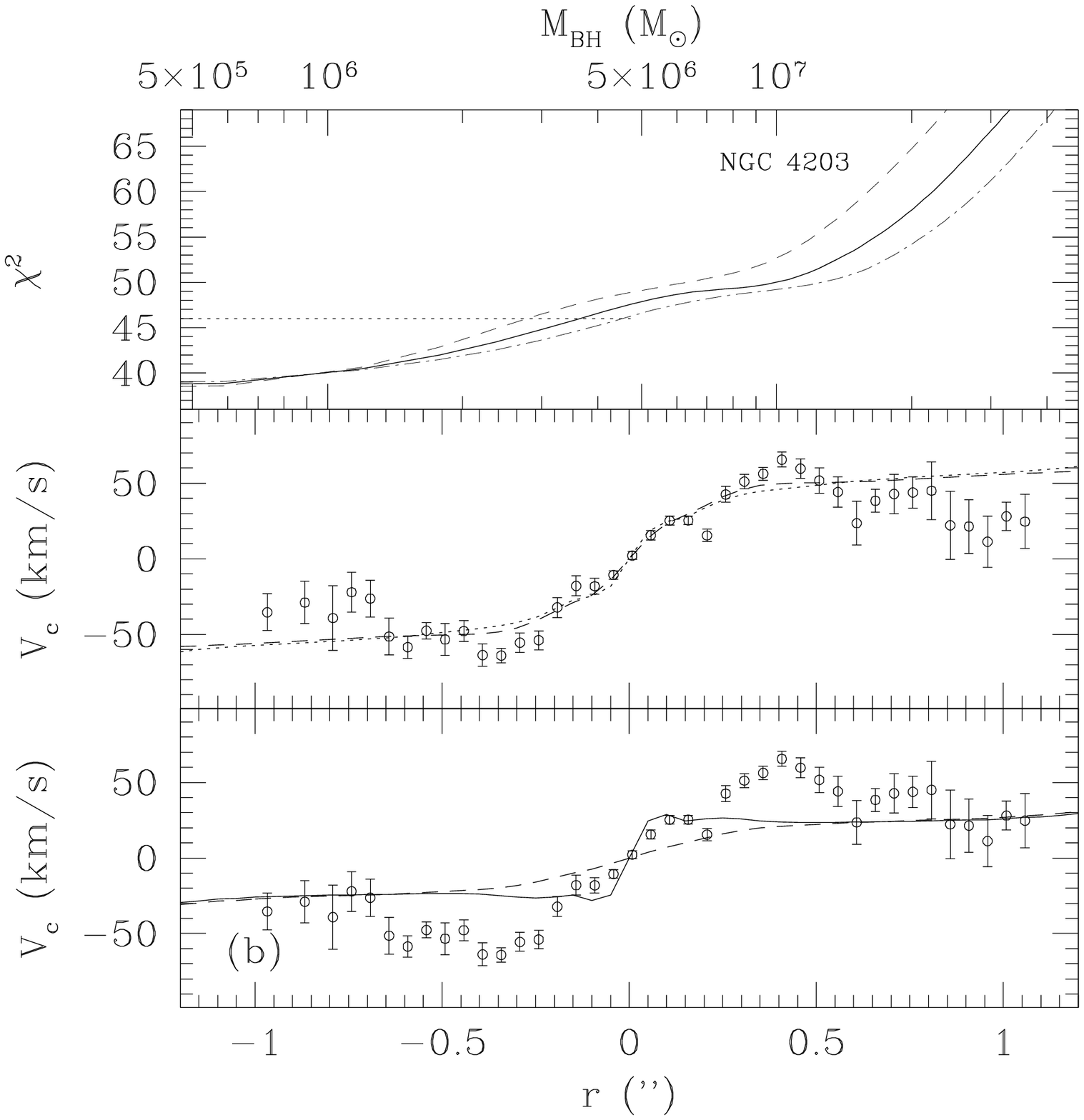}
\end{figure}
\begin{figure}
\plotone{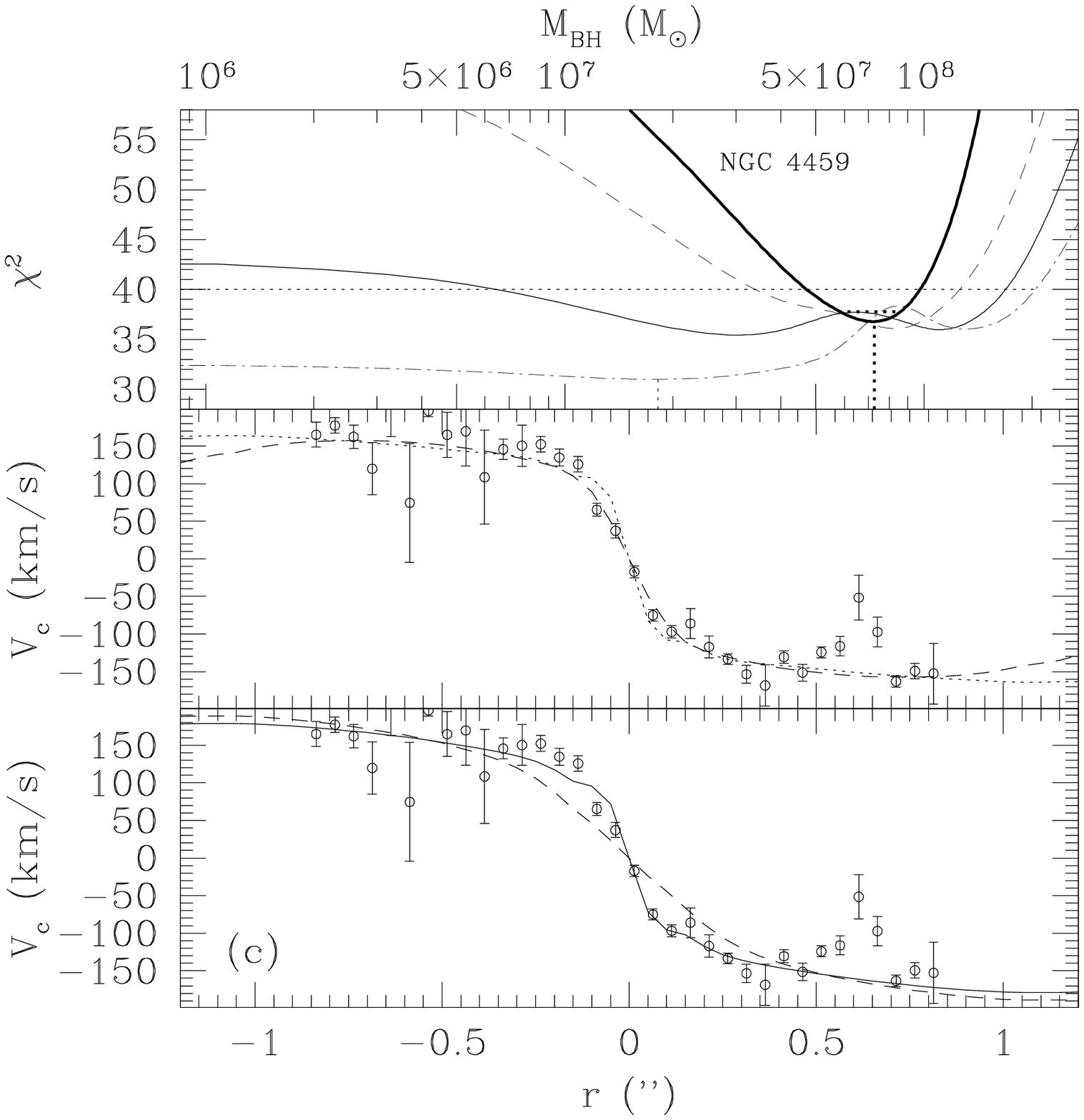}
\end{figure}
\begin{figure}
\plotone{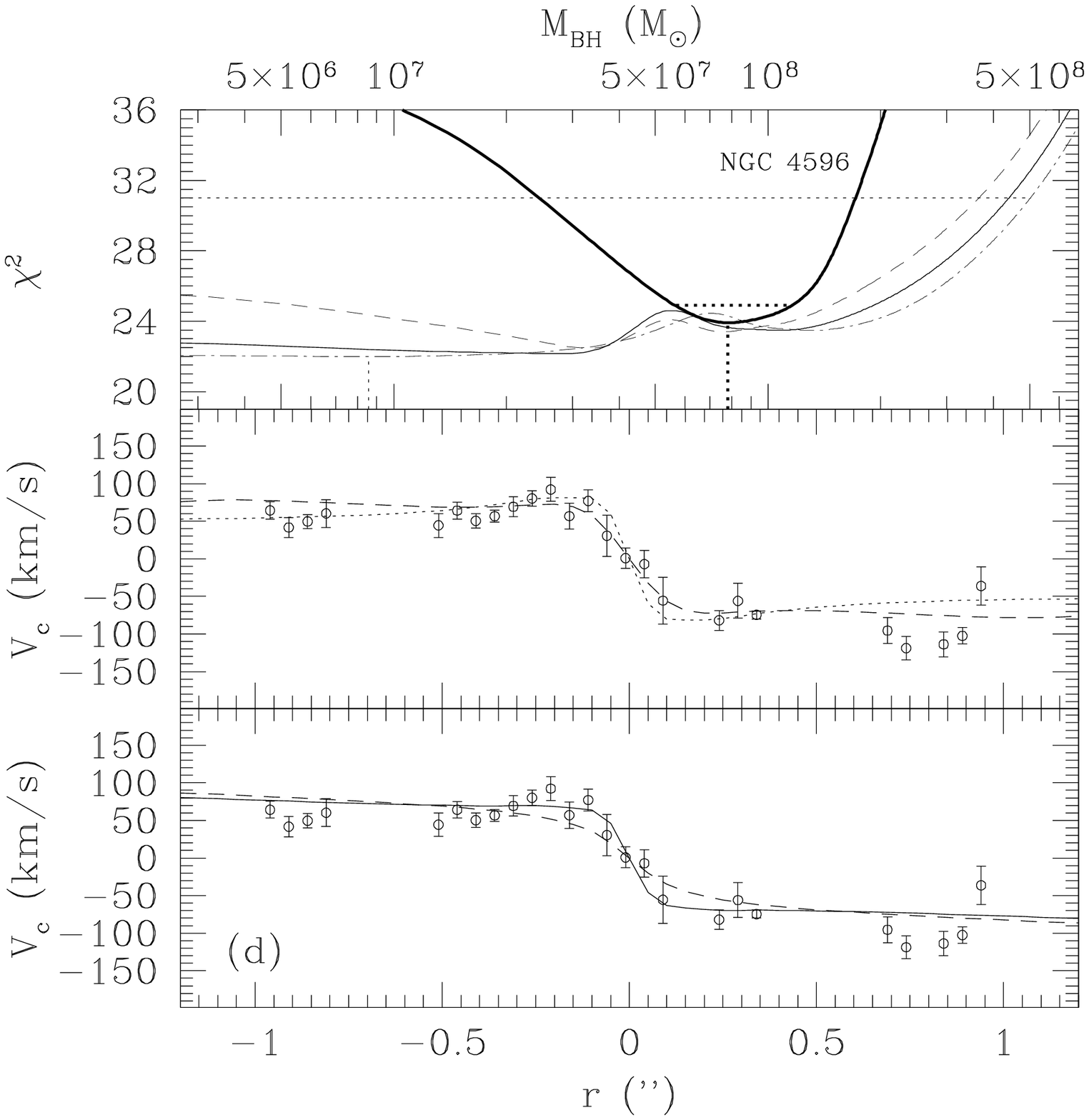}
\end{figure}
 
\begin{figure}
\plotone{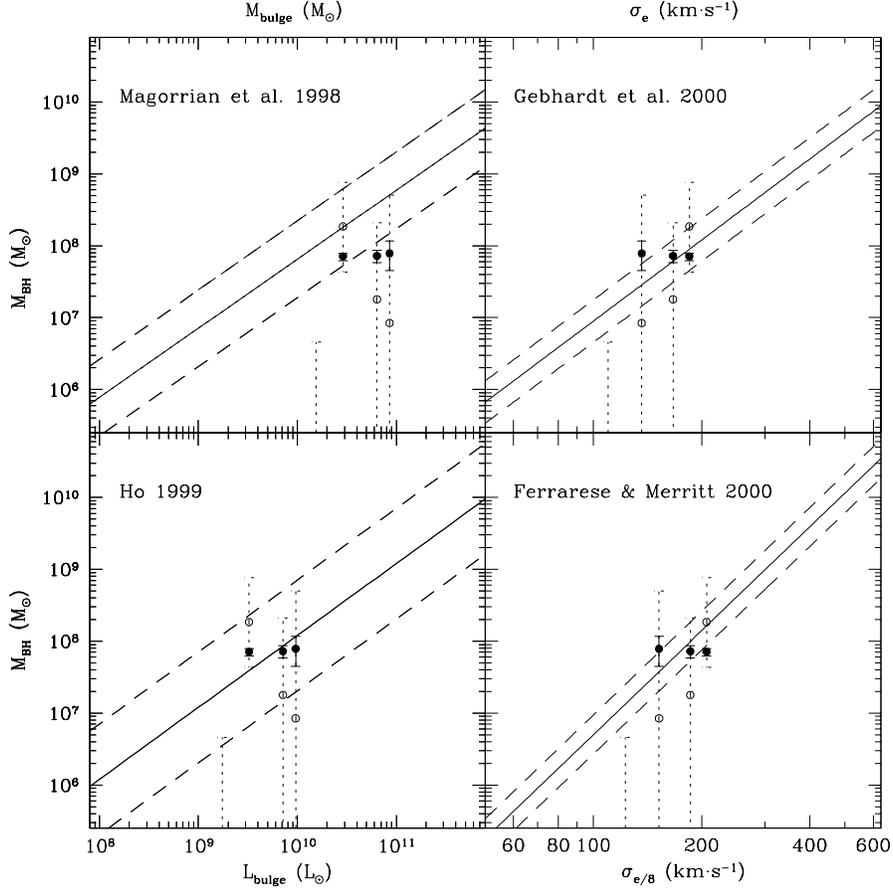}
\caption{$M_\bullet$ versus $M_{bulge}$ and $L_{bulge}$ ({\sl left panels}), 
and versus $\sigma_e$ and $\sigma_{e/8}$({\sl right panels}).  The solid
diagonal lines represent the correlation fits reported previously by
the indicated authors, with dashed diagonal lines representing the
$1\sigma$ scatter in $M_\bullet$. Best black-hole mass estimates or upper 
limits are shown for models with disk unconstrained orientation 
(open circles, dashed limits) and constrained orientation (filled circles, solid limits). 
Error bars correspond to $1\sigma$ and $3\sigma$ confidence limits, 
in the constrained and unconstrained case respectively.}
\label{fig:SMBH_in_bulges_fig5}  
\end{figure}

\end{document}